\documentclass[manuscript,screen,nonacm]{acmart}
\acmConference[]{} 

\usepackage{subcaption}
\usepackage{multirow}
\usepackage{xcolor}
\usepackage{colortbl}
\usepackage{mathtools}

\usepackage{pifont}

\usepackage{wrapfig}

\usepackage{array}
\newcolumntype{C}[1]{>{\centering\let\newline\\\arraybackslash\hspace{0pt}}m{#1}}

\newcommand{\workshopname}{GenAICHI: CHI 2024 Workshop on Generative AI and HCI}
\newcommand{\licensedetails}{Licensed under a Creative Commons Attribution 4.0 International License (CC BY 4.0). Copyright remains with the author(s).}
\newcommand\extrafootertext[1]{
    \bgroup
    \renewcommand\thefootnote{\fnsymbol{footnote}}%
    \renewcommand\thempfootnote{\fnsymbol{mpfootnote}}%
    \footnotetext[0]{#1}%
    \egroup
}

\AtBeginDocument{ 
    \fancypagestyle{firstpagestyle}{
        \fancyhf{}
        \fancyfoot[L]{\sffamily\footnotesize \workshopname}%
        \fancyfoot[C]{\sffamily\footnotesize \thepage}
    }
    \fancyhf{}
    \fancyhead[L]{\sffamily\footnotesize\shorttitle}
    \fancyhead[R]{\sffamily\footnotesize\shortauthors}
    \fancyfoot[L]{\sffamily\footnotesize\workshopname}%
    \fancyfoot[C]{\sffamily\footnotesize\thepage}
    \extrafootertext{\licensedetails}
}
\usepackage{lipsum} 

\begin{document}

\graphicspath{{figures/}}

\title{Interaction Design for Human-AI Choreography Co-creation}

\author{Yimeng Liu}
\affiliation{
  \institution{University of California, Santa Barbara}
  \city{Santa Barbara}
  \country{USA}}
\email{yimengliu@cs.ucsb.edu}

\begin{abstract}
Human-AI co-creation aims to combine human and AI strengths for artistic results exceeding individual capabilities. Frameworks exist for painting, music, and poetry, but choreography's embodied nature demands a dedicated approach. This paper explores AI-assisted choreography techniques (e.g., generative ideation, embodied improvisation) and analyzes interaction design --- how humans and AI collaborate and communicate --- to inform the design considerations of future human-AI choreography co-creation systems.
\end{abstract}



\maketitle

\section{Introduction}
Human-AI co-creativity is a collaborative process where humans and AI work together as partners to create innovative solutions, artistic works, or other creative outputs. This process depends heavily on the interaction dynamics, roles of each participant, and communication styles employed. Careful design of these elements is essential for maximizing the effectiveness and benefits of human-AI co-creative systems, such as increased efficiency and enhanced creativity. 
Building upon an established interaction design framework for human-AI co-creativity~\cite{rezwana2023designing} across domains like painting, music, storytelling, and poetry~\cite{white2017generating, lopez2010real, osone2021buncho, spoto2024library}, this paper focuses on a relatively unexplored domain: choreography co-creation with AI. This inherently embodied and highly creative research field needs tailored interaction design insights to unlock its full potential. 
In this work, we present existing AI-supported choreography systems and techniques and analyze their interaction design through the lens of three distinct design goals: \textit{choreography generation}, \textit{creativity support}, and \textit{human-AI choreography co-creation}. Inspired by the computational creativity research by Davis et al.~\cite{davis2015enactive}, we categorize existing systems based on these goals and uncover three key interaction design considerations: facilitating parallel and spontaneous interaction between humans and AI, assigning distinct yet complementary roles to humans and AI, and ensuring effective human-AI communication. These insights aim to serve as a resource for designing future systems and refining existing ones, ultimately pushing the boundaries of human-AI co-creation in choreography.

\section{Related Work} \label{sec:related_work}
\subsection{Choreography Generation}
Previous research on AI-assisted choreography generation has primarily aimed at developing techniques for automatically creating innovative, unexpected, and valuable dance concepts and materials. Much of this work has explored using generative AI models, such as diffusion models, to facilitate this process. These models have taken diverse input modalities like music, text, and video, transforming extracted features into dance movements~\cite{alexanderson2023listen, tseng2023edge, chuang2022music2dance, zhang2022music, ye2020choreonet, gong2023tm2d, chan2019everybody}.

\subsection{Creativity Support}
Prior work on AI-based creativity support has utilized techniques like tracking history, simulating possibilities, and exploring alternatives to assist individuals in their creative endeavors. For instance, systems like~\cite{crnkovicfriis2016generative, pettee2019imitation, liu2024exploring} have leveraged generative AI to augment creativity during choreography ideation. These systems have empowered users to generate new movements, iteratively edit dance sequences, document creative practice, and foster the exploration of both system-generated and user-provided ideas, thereby supporting user creative potential. 

\subsection{Human-AI Choreography Co-creation}
Research on human-AI choreography co-creation has worked on developing co-creative agents that engage in real-time improvisation with humans to enrich the creative process. Systems like Viewpoints AI~\cite{jacob2013viewpoints}, LuminAI~\cite{long2017designing}, and Robodanza~\cite{infantino2016robodanza} have fostered collaborative engagement during choreography. These systems have enabled humans and AI to take spontaneous initiatives, contributing jointly to the creation of dance movements.

\section{Interaction Design for AI-supported Choreography Creation} \label{sec:interaction_design}
Building on Ciolfi et al.'s four-stage choreography creation process~\cite{ciolfi2016choreographers}: \textit{preparation}, \textit{studio}, \textit{performance}, and \textit{reflection}, we focus on the first two stages where AI can shine. Leveraging its content generation capabilities, AI can empower choreographers during ideation and prototyping. Drawing on relevant research for each stage, we analyze human-AI interaction through the lens of Rezwania et al.'s co-creative framework for interaction design (COFI)~\cite{rezwana2023designing}, including the \textit{collaboration} and \textit{communication} styles between humans and AI. 

\subsection{Interaction Design for Choreography Ideation in the Preparation Stage}
The preparation stage focuses on ideation and crafting choreographic materials. However, research on human-AI co-creation for this stage remains scarce, so we discuss techniques developed for choreography generation and creativity support and explore how these techniques can be expanded to foster co-creativity regarding interaction design. 

\subsubsection{Collaboration Style}
Most previous research has adopted a turn-taking collaboration style in the preparation stage, where humans and AI alternate to contribute to the same or separate tasks. In the same-task scenario, humans utilize AI-based techniques to generate artifacts with convergent or divergent ideas. For example, existing work~\cite{crnkovicfriis2016generative, gong2023tm2d, liu2024exploring} allows both humans and AI to contribute to the same dance sequences. The underlying generative AI model is called upon when humans initiate dance generation or modification. Conversely, AI-based methods can potentially support the evaluation of created artifacts or user-provided concepts in divided tasks. This branch has not been fully explored in prior work. However, leveraging effective human motion evaluation techniques, such as Laban Movement Analysis~\cite{laban1975modern}, can enhance the understanding of abstract movements and contribute to choreography creation. Regarding the timing of initiative, AI typically responds to human requests when ideation or evaluation is needed, as prior research has shown that users tend to be opposed to AI taking the lead in turn-taking interaction~\cite{winston2017turn}. 

\subsubsection{Communication Style}
During the preparation stage, where intense brainstorming is key, an on-demand interaction design is necessary to balance creative thinking with the absorption of new information. To achieve this, clear and direct communication between humans and AI is important. Human-to-AI communication can leverage intuitive methods like text, voice, and direct manipulation. These methods can facilitate the seamless transmission of needs and creative vision, as demonstrated by choreographers who utilize them to communicate their ideas effectively~\cite{ciolfi2016choreographers, zhou2021dance}. AI-to-human communication can rely on easily understandable text and visuals to present dance poses or sequences, as well as evaluation results for movements. 

While intentional communication plays a key role, few efforts have explored the potential of consequential communication in human-AI interaction. This approach complements intentional communication when direct conversation fails to capture a nuanced creative vision. For example, choreographers often struggle to articulate implicit feelings in dance ideas through words alone~\cite{rezwana2023designing}. They may rely on sound, facial expressions, or gestures to convey these nuances. This presents a challenge for AI in processing such subtle information and offering relevant choreographic materials that align with the artist's intent. Furthermore, research on mixed-initiative human-AI communication for co-creativity is limited. Understanding how the level of interaction, e.g., reactive vs. proactive AI, impacts human-AI communication in the preparation stage remains an open question. 

\subsection{Interaction Design for Choreography Prototyping in the Studio Stage}
Shifting gears to the studio stage, the focus is on translating ideas into movement and collaborating with other dancers and choreographers. Here, embodiment becomes essential in interaction design. By analyzing the interaction design of existing co-creativity systems in this stage, we uncover current challenges and pose open research questions.

\subsubsection{Collaboration Style}
Existing research has explored parallel collaboration styles, where humans and AI share mixed initiatives to contribute to a shared task. Examples include Viewpoints AI~\cite{jacob2013viewpoints}, LuminAI~\cite{long2017designing}, and Robodanza~\cite{infantino2016robodanza}, all designed to facilitate real-time, collaborative dance improvisation and performance. These systems have enabled AI to capture and process human motion and generate new dance movements embodied by projections or robots that complement or react to the human dancers. Importantly, initiative timing is spontaneous, with humans and AI free to initiate and modify dance poses and movements, contributing to the evolving artifact.

\subsubsection{Communication Style}
For an unobtrusive and immersive experience in the studio stage, interaction design requires mirroring human communication styles through both explicit and implicit methods. Human-to-AI communication can utilize intentional methods like voice and direct manipulation alongside consequential methods like facial expressions and embodied cues. This aligns with how humans naturally communicate, offering a broader spectrum of information exchange. AI, on the other hand, can utilize speech, haptics, and visuals to respond. 

Previous research often overlooks the design of human-to-AI consequential and AI-to-human communication in the studio stage despite their crucial role in fostering embodied experiences. Just like humans observing others to understand their movement and intent, AI needs to develop a similar Theory of Mind~\cite{premack1978does} to interpret human mental states beyond explicit instructions. In the studio stage, where information exchange is frequent and initiative is spontaneous, relying solely on explicit communication hinders AI's effectiveness as a collaborator and communicator. Consequently, AI systems need to be proactive and sensitive to implicit information to achieve true collaboration.

\section{Discussion and Future Directions}

\begin{table}[!ht]
    \centering
    \caption{Design and interaction of choreography-support systems in the choreography preparation and studio stages.}
    \scalebox{0.77}{
    \begin{tabular}{p{0.17\textwidth}p{0.09\textwidth}p{0.22\textwidth}p{0.12\textwidth}C{0.206\textwidth}p{0.12\textwidth}C{0.206\textwidth}}
    \toprule
        \multirow{2.5}{*}{\textbf{Type}} & \multirow{2.5}{*}{\textbf{Paper}} & \multirow{2.5}{*}{\textbf{Description}} & \multicolumn{2}{c}{\textbf{Preparation Stage}} & \multicolumn{2}{c}{\textbf{Studio Stage}} \\
         \cmidrule(r){4-5}\cmidrule(l){6-7}
         &  &  & \textbf{Collaboration} & \textbf{Communication} & \textbf{Collaboration} & \textbf{Communication} \\
         \midrule
        Choreography Generation & \cite{alexanderson2023listen, tseng2023edge, chuang2022music2dance, zhang2022music, ye2020choreonet, gong2023tm2d, chan2019everybody} & Convert multimodal input into dance motion & Turn-taking, shared task, reactive & \begin{tabular}{@{}p{0.207\textwidth}@{}} Human-->AI: Intentional AI-->Human: Intentional \end{tabular} & \cellcolor{gray!10} & \cellcolor{gray!10} \\
        \midrule
        Creativity Support & \cite{crnkovicfriis2016generative, pettee2019imitation, liu2024exploring} & Augment creativity via interaction with the system & Turn-taking, shared task, reactive & \begin{tabular}{@{}p{0.207\textwidth}@{}} Human-->AI: Intentional AI-->Human: Intentional \end{tabular} & \cellcolor{gray!10} & \cellcolor{gray!10} \\
        \midrule
        Human-AI Choreography Co-creation & \cite{jacob2013viewpoints, long2017designing, infantino2016robodanza} & Co-create dance based on collaborative engagement & \cellcolor{gray!10} & \cellcolor{gray!10} & Parallel, shared task, proactive & \begin{tabular}{@{}p{0.207\textwidth}@{}} Human-->AI: Intentional AI-->Human: N/A \end{tabular} \\
    \bottomrule
    \end{tabular}
    }
    \Description{}
    \label{tab:rw_comparison}
\end{table}

\begin{wrapfigure}{r}{0.448\textwidth}
    \includegraphics[width=\linewidth]{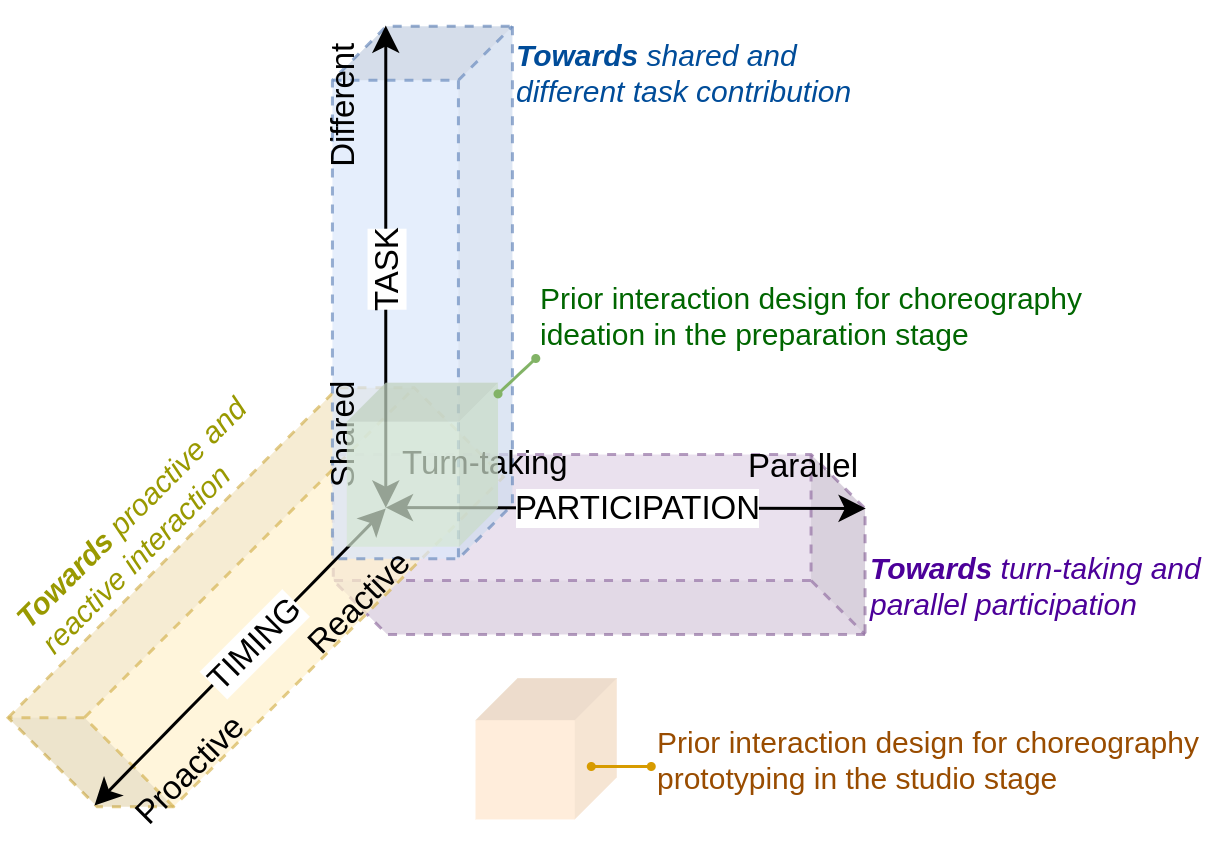} 
    \caption{Interaction design space for AI-supported choreography creation. The three axes are built upon the co-creative framework for interaction design introduced in ~\cite{rezwana2023designing}.}
    \label{fig:interaction_design_space}
\end{wrapfigure}

Table~\ref{tab:rw_comparison} summarizes the discussed research, comparing their design and interaction approaches covered in Sections~\ref{sec:related_work} and~\ref{sec:interaction_design}. This analysis yielded three key insights for designing future human-AI choreography co-creation systems. We leverage these insights to explore the interaction design space in Figure~\ref{fig:interaction_design_space}, which incorporates factors like participation style, task distribution, and initiative timing (inspired by Rezwan et al.~\cite{rezwana2023designing}). 

\subsection{Building Parallel and Spontaneous Collaboration}
Previous research focuses on parallel and spontaneous human-AI collaboration in the studio stage, neglecting the potential for AI to be a true partner throughout the entire process, including in the preparation stage. Specifically, AI systems usually wait their turn to assist in brainstorming and refining ideas, often when humans request them. This turn-taking style positions them as tools rather than collaborators, as effective collaboration thrives on spontaneous exchange of feedback, which is crucial for successful communication and task completion. Therefore, future research can focus on developing AI that transcends simply waiting for its turn. By actively engaging in the creative process, these AI systems could significantly enhance collaboration. Imagine AI that offers timely inspiration, provides constructive feedback, and proposes refinements throughout choreography creation --- acting as a concurrent source of creative input, independent of human work at times. This shift would foster a more dynamic and collaborative experience. 

\subsection{Designing Complementary Roles for Human and AI}
Existing research on human-AI choreography collaboration often overlooks the crucial aspect of task division. While creating new choreography gets ample focus, tasks like expanding, refining, and transforming existing pieces remain largely unexplored. This gap might be due to current AI systems often mimicking user input or existing works, limiting their ability to generate truly innovative and thought-provoking pieces. However, such outputs can be instrumental in sparking divergent thinking, a technique proven to enhance creativity~\cite{liu2024exploring}. In essence, these AI-generated pieces could ignite a deeper exploration of creative concepts and materials, offering a wider range of possibilities to build upon existing choreography~\cite{kantosalo2016modes}. 

\subsection{Enabling Effective Dialogue and Mutual Understanding}
The reviewed papers highlight a gap in effective human-AI communication during choreography creation. In the preparation stage, the interaction leans heavily towards a one-way flow. Humans initiate ideas and instructions, while the AI passively responds. The dynamic improves somewhat during the studio stage, where both humans and AI contribute elements to the dance piece. However, achieving direct communication from AI to humans similar to human-to-human interaction remains challenging.
Moving forward, research can explore the potential of proactive AI communication styles. Imagine an AI system that actively monitors a dancer's movements and offers constructive suggestions based on its observation. This would mirror the dynamic of a human collaborator, fostering a richer creative process. Furthermore, integrating consequential communication from humans to AI is crucial to fostering a more natural and immersive co-creative experience. This is especially important in the highly embodied realm of choreography.

\bibliographystyle{ACM-Reference-Format}
\bibliography{chi2024_workshop}

\end{document}